\pdfoutput=1

\documentclass[pra,10pt,superscriptaddress, footnoteinbib]{revtex4}
\usepackage{amsmath}
\usepackage{latexsym}
\usepackage{amssymb}
\usepackage{graphics,epstopdf}
\usepackage[colorlinks=true, citecolor=blue, urlcolor=blue ]{hyperref}
\usepackage{epsf,graphics,graphicx}

\newcommand{\ed}{\end{document}}
\newcommand{\beq}{\begin{equation}}
\newcommand{\eeq}{\end{equation}}

\begin{document}
	\title{Spin Transport in Tilted Electron Vortex Beams}
	\author{ Banasri Basu\footnote{Electronic address:{sribbasu@gmail.com}}${}^{}$, Debashree Chowdhury\footnote{Electronic
			address:{debashreephys@gmail.com}  }}
	\affiliation{Physics and
		Applied Mathematics Unit, Indian Statistical Institute,\\
		203,Barrackpore Trunk Road, Kolkata 700 108, India}
\begin{abstract}
 In this paper we have enlightened the spin related issues of tilted Electron vortex beams. We have shown that in the skyrmionic model of electron we can have the spin Hall current considering the tilted type of electron vortex beam. We have considered the monopole charge of the tilted vortex as time dependent and through the time variation of the monopole charge we can explain the spin Hall effect of electron vortex beams. Besides, with an external magnetic field we can have a spin filter configuration. 
\end{abstract}
\maketitle


\section{Introduction}

 Vortex beams can be defined as a beam of particles either of photons or any other particles that are freely propagating and have a special kind of property i.e quantized orbital angular momentum (OAM). Propagating waves carrying intrinsic orbital angular momentum (OAM) are a widely extained research area in condensed matter physics. In optical vortices, light is twisted like a corkscrew around its axis of travel and when the whole picture is projected on a flat surface what we have is the darkness at the center, commonly known as phase singularity. The number of the twist determines the rate of the spinning of light around the axis and spinning carries OAM. Light waves carry two types of OAM (External and Internal)\\
  $ \longrightarrow $ The external one: $\vec{L} = \vec{r}\times \vec{p} ,$ which is origin dependent.\\
  $ \longrightarrow $ The internal OAM, which is origin-independent angular momentum of light beams and associated with a helical or twisted wavefront.  In search of a vortex beam of matter people try to find whether it is possible to have electron vortex beams (EVB)? Because of the fact that electron wavelength is much smaller than that of visible light,  it is possible to produce much better images in an electron microscope to be taken than currently possible. Inspired by the work of Nye and Berry (1974) about the helical wave fronts and phase singularities in radio waves, the theoretical description of EVB is put forward by Bliokh et al \cite{4,5}. Later in 2010 Uchida and Tonomura  \cite{1}produced EVB using spiral phase plate and also Verbeeck et al produced it with holographic reconstruction method \cite{2}. These EVB can be visualized as scalar electrons orbiting around vortex lines. So three situations can arise in this connection\\
  
   1. Vortex line parallel (2. orthogonal) to the wave propagation direction are termed as screw(edge)dislocation respectively. \\
     3. Mixed screw-edge dislocation is the situation where vortex lines are tilted with respect to the propagation direction.  
 Bliokh et. al (2012) show that for a mixed screw-edge dislocation having tilted vortex lines carry well defined OAM in an arbitrary direction. This bears a close analogy between optical and matter waves. Our goal is to study the spin properties of tilted EVB from the geometric phase aspect. 
 
 We have considered here the skyrmionic model of electron in which a scalar particle encircling a vortex line is considered to be topologically equivalent to a scalar particle encircling a magnetic flux line. The quantization of a fermion can be achieved when we introduce a direction vector at each space-time point which appears as a vortex line.
 
 Furthermore, what happens if one considers the external magnetic field in the longitudinal direction \cite{14} is also discussed in this work. Here interestingly we have clustering of spins and spin filtering. One can also be able to have alternating spin current for a hopping field.
 
 The organization of the field is as follows: In section II we focus on the skyrmionic model of electron. This follows with section III with the spin Hall effect of electron vortex beams. In section IV we have discussed the effect of using a longitudinal magnetic field in the direction of propagation of the EVBs. We have end our discussion in the section V. 

\section{skyrmionic model}
Bandyopadhyay and Hajra in some of their papers \cite{7,8} demonstrated that the quantization of a fermion can be achieved when we introduce an internal variable that appears as a direction vector at each and every space time points in Nelson's stochastic quantization procedure (Nelson  1966, 1967) \cite{9,10}. This direction vector essentially corresponds to the spin degree of freedom. Basically it is a SL(2,C) gauge theory and by demanding hermiticity we can have gauge fields belonging to the unitary group SU(2).
In this skyrmionic model \cite{11,12} of electron we can write the space-time coordinate and momentum as
{\begin{eqnarray}\label{eq1}
 Q_\mu &= & -i\left( \frac{\partial}{\partial p_\mu}+{\cal{A}}_\mu (p)\right) \nonumber \\
 P_\mu &= & i\left( \frac{\partial}{\partial q_\mu}+{\cal{B}}_\mu (q)\right)
 \end{eqnarray}
 where ${\cal{A}}_\mu ({\cal{B}}_\mu)$ is the momentum(spatial coordinate)dependent SU(2) gauge field. This leads to the noncommutative properties of the coordinate and momentum as
 \begin{eqnarray}\label{eq2}
 [Q_{\mu}, Q_{\nu}] =  {\cal{F}}_{\mu \nu}(p),~~~~
 [P_{\mu}, P_{\nu}] =  {\cal{F}}_{\mu \nu}(q),
 \end{eqnarray}
 where \begin{equation} \label{eq3}
 {\cal{F}}_{\mu \nu}=\partial_\mu {\cal {A}}_\nu- \partial_\nu{\cal {A}}_\mu +[{\cal {A}}_\mu, {\cal {A}}_\nu] 
 \end{equation} is the curvature and its
 functional dependence  represents the existence of monopoles in real space \cite{16, 17} and we can write $
 [p_i,p_j] =i \mu \epsilon_{ijk}\frac{x_k}{r^3}\label{eq4}
 ,$
 $\mu$ is the monopole strength.
 \begin{figure}
   \includegraphics[width=7.0 cm]{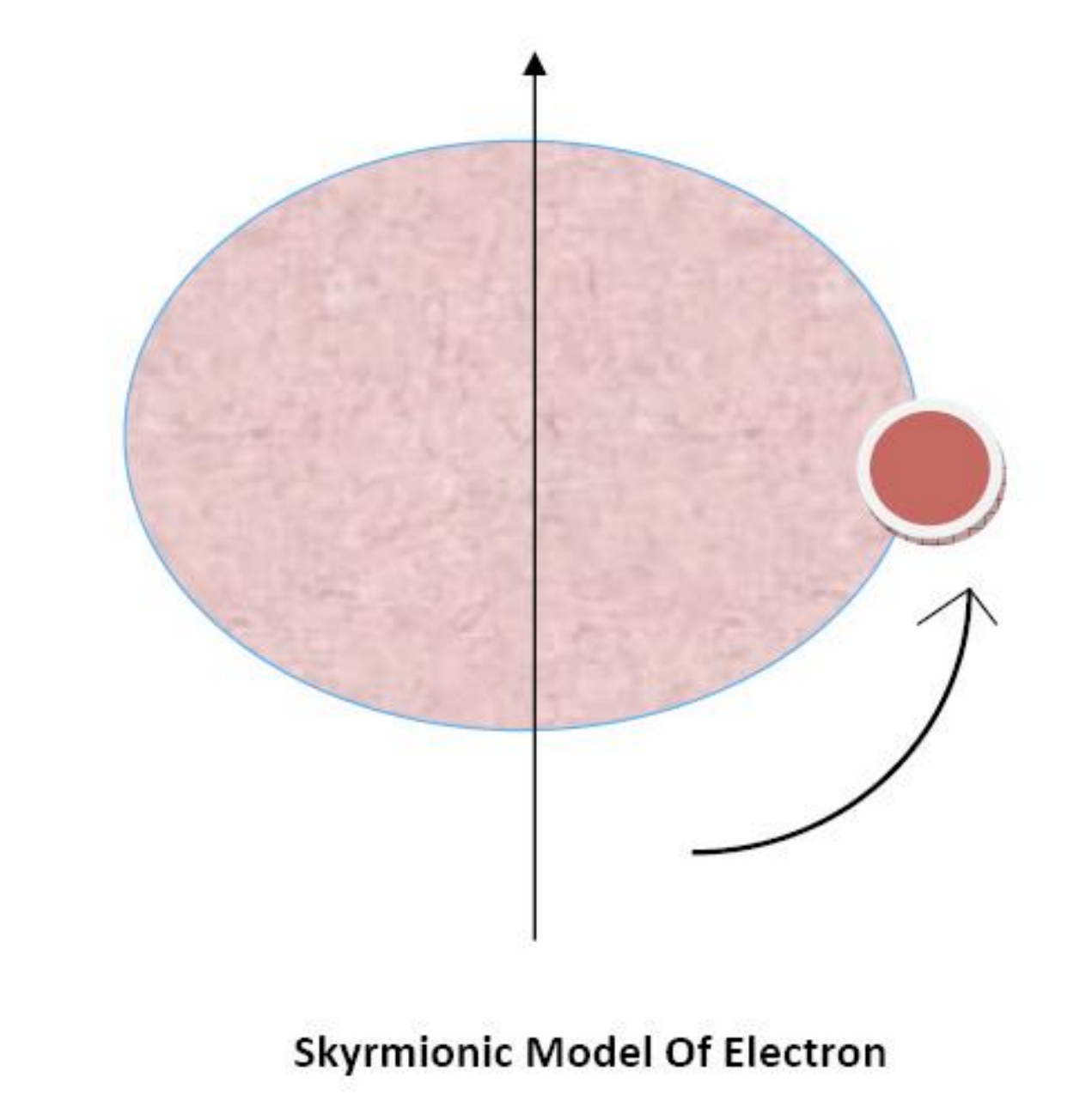}
   \caption{\label{a} Skyrmionic model of electron.}
   \end{figure}
When the scalar particle encircles a loop enclosing magnetic flux lines the associated Berry phase \cite{18} is given by \begin{equation} \phi_B=2\pi \mu \end{equation} and $\mu=1/2$ corresponds to one magnetic flux line. Here a fermion may be visualized as a scalar particle encircling a magnetic flux line. In this skyrmionic picture of an electron  EVB  is a natural consequence. Scalar electron encircling the vortex line can carry quantized OAM. When the monopole is situated at the origin of a unit sphere, the Berry phase can be written as $\phi_{B} = \mu\Omega(C),$ where $\Omega(C),$ is the solid angle at the origin  and can be written as 
\begin{equation}  \Omega(C) = \int_{C}(1 - cos \theta)d\phi = 2\pi(1- cos\theta).\label{eq8}\end{equation}
From equations (4) and (5) we can write the effective monopole charge associated with a vortex line, having polar angle $\theta$ with the $z$ axis as
\begin{equation} \mu = \frac{1}{2}(1 - cos\theta)\label{eq12}.\end{equation}
For $\theta = 0$ and $\frac{\pi}{2}$ it takes quantized values but for other angles $0 < ~\theta<~\frac{\pi}{2},$  $\mu$ is non-quantized.

Thus if we consider this model we can interpret electron as a scalar particle encircling a vortex line in a specific direction, the solution of the Dirac equation in cylindrical coordinates exhibits the vortex-dependent properties explicitly. Most importantly the SOI factor($\Delta$), which plays an important role in the spin transport phenomena is determined by the Berry phase. In relativistic limit one can show $\Delta = sin^{2}\theta = 4\mu(\mu -1).$ For $\mu = 0(1),$ corresponding to $\theta = 0(\pi),$ we have $\Delta = 0$ which indicates that the vortex line is parallel (anti-parallel) to the $z$ axis and SOI vanishes. $\Delta$ corresponds to the non-quantized monopole charge for other values of $\theta.$ The SOI term has been incorporated through the gauge theoretic methods by several authors. This effectively leads to the interaction of the spin with a magnetic field in momentum space. We also proceed in the same direction and can write the generalized OAM as 
$ \vec{\tilde{L}} = \vec{Q} \times \vec{p} = \vec{q}\times \vec{p} + {\cal \vec{A}}(\vec{p})\times \vec{p} = \vec{L} + \vec{L}^{'} \label{21},$
with $ \vec{L}^{~'} =  {\cal \vec{A}}(\vec{p})\times \vec{p}\label{22}.$
From above equation we can write
$ \vec{L}^{~'} = -\mu \vec{\kappa}\times (\vec{\kappa}\times \vec{\sigma})\label{24}.$
The expectation value of $\vec{\sigma}$ i.e $\langle\vec{\sigma}\rangle$ is given by
$ \langle\vec{\sigma}\rangle = \frac{\langle\psi|\vec{\sigma}|\psi\rangle}{\langle\psi|\psi\rangle}\label{2},$
where $\psi$ is a two-component spinor
$ \psi = \left(\begin{array}{cr}
   \psi_{1} \\
    \psi_{2}
 \end{array}\right) \label{222}$ with $\langle\psi|\psi\rangle = 1.$
 This gives $ \langle\vec{\sigma}\rangle = \langle\psi|\vec{\sigma}|\psi\rangle = \vec{n}\label{28},$
 with $\vec{n}^{2} = 1.$
 We can write
$ \langle\vec{L}^{'}\rangle = -\mu \vec{n}\label{29}.$
 Here $ \vec{\kappa} = \frac{\vec{p}}{p}.$ Considering the total angular momentum to be constant we can write \cite{6} 
  \begin{eqnarray}\label{100}
  \langle\vec{\tilde{L}}\rangle &=& (l - \mu)\hat{\vec{z}}\nonumber\\
  \langle\vec{\tilde{S}}\rangle &=& (s + \mu)\hat{\vec{z}},
  \end{eqnarray}
 Importantly, a part of the spin angular momentum is converted to OAM, which practically implies the SOI effect.
 This SOI is an important concept in spin Hall phenomenon. In the next section we will discuss the SHE of electron vortex beams.
  \section{Spin Hall effect in Tilted Electron vortex beam}
The monopole charge in $3+1$ dimensions is equivalent to the central charge $c$ of the conformal field theory in $1+1$ dimensions \cite{pb}. Thus if the central charge $c$ undergoes the renormalization group (RG) flow, the monopole charge $\mu$ also undergoes the same. Which indicates that when $\mu$ depends on a certain parameter $\lambda$ we have\\

  - $\mu$ is stationary at fixed points $\lambda^{*}$ of the RG flow.\\
  - at the fixed points $\mu(\lambda^{*})$ is equal to the monopole charge $\mu$ given by the quantized values $0$, $\pm\frac{1}{2},$ $\pm 1$....\\
  -$\mu$ decreases along the RG flow i.e $L\frac{\partial \mu}{\partial L}\leq 0,$ where $L$ is a length scale. \\
  As we can write $(L = ct)$ we can write
 \begin{eqnarray} t\frac{\partial\mu}{\partial t}\leq 0\label{32},\end{eqnarray} which implies that we can take $\mu$ as the time dependent parameter. In fact when $\mu$ takes a value $\tilde{\mu}(t)$ on the RG flow which is non-quantized and at certain fixed value of time it takes the quantized value $\mu.$ For explicit time dependence of the monopole charge, $\tilde{\mu}(t),$ we have the corresponding gauge field explicitly time dependent. An electric field $\vec{{\cal E}} = \frac{\partial\vec{A}}{\partial t}$ is now generated.
 This electric field accelerates electrons so that the momentum carries explicit time dependence.
 In fact the time dependence of $\tilde{\mu}$ is here incorporated through the time dependence of momentum, denoted by $\vec{k}.$  Here we can consider a non-inertial coordinate frame with basis vectors $(\vec{v}, \vec{w}, \vec{u})$ attached to the local direction of momentum $\vec{u} = \frac{\vec{k}}{k}.$ This coordinate frame rotates with respect to a laboratory frame with some angular velocity \cite{6}. Now taking the direction of the vortex line at an instant of time as the local $z$ axis which represents the direction of propagation of the wave front, we have the paraxial beam in the local frame. 
 Now if the direction of the vortex line is taken to be the local $z$ axis in the non-inertial frame, the local value of $\tilde{\mu}$ is changed to the quantized value $|\mu| = \frac{1}{2}$ due to the precession of the spin vector.  We can write the corresponding Berry curvature as
$ \vec{\Omega}(\vec{k}) = \mu\frac{\vec{k}}{k^{3}}.\label{40}$ This curvature will give rise to an anomalous velocity given by
\begin{eqnarray}\vec{v}_{a} = \dot{\vec{k}}\times \vec{\Omega}(\vec{k}) = \mu \dot{\vec{k}}\times\frac{\vec{k}}{k^{3}}. \label{41}\end{eqnarray}
This anomalous velocity gives rise to the spin Hall effect. 
Thus the anomalous velocity points along opposite directions depending on the chirality  $s_{z} = \pm \frac{1}{2}$ corresponding to $\mu> 0(< 0).$ This separation of the spins gives rise to the spin Hall effect (Dyakonov and Perel 1971). Thus a tilted vortex in the inertial frame carrying OAM will give rise to relativistic spin Hall effect. The expression of spin current can be written as 
\begin{equation}
 j^{k}_{s,j} = \frac{\hbar}{4}\left\langle \{\sigma^{k}, v_{j}\}\right\rangle.\end{equation}
This is a very important expression as it suggests us that the spin current depends on the parameter of monopole strength $\mu,$ when we substitute eqn.(9) in the above equation. Next we focus on the study of using an external magnetic field along the propagation direction of electron vortex beam.
   \section{electron vortex beams in a magnetic field}
   If the magnetic field be axially symmetric longitudinal one, the vortex vector potential can be chosen as \cite{14} $ \vec{A}({\vec r}) = \frac{B(r)r}{2}\hat{\vec{e}}_{\phi},$ where $\vec{B} = \vec{\nabla}\times\vec{A}.$ In the sharp point limit the Schrodinger equation in cylindrical coordinates, in terms of the external observable variables, is given by \cite{14}
\begin{equation} -\frac{\hbar^{2}}{2m}\left[\frac{1}{r}\frac{\partial}{\partial r}\left(r\frac{\partial}{\partial r}\right) + \frac{1}{r^{2}}\left(\frac{\partial}{\partial \phi} + ig\frac{2r^{2}}{w_{m}^{2}}\right)^{2} + \frac{\partial^{2}}{\partial z^{2}}\right]\psi = E\psi,\label{4}\end{equation}
   where $w_{m} = 2\sqrt{\frac{\hbar}{|eB|}}$ is the magnetic length parameter and $g = sgn B = \pm 1$ indicates the direction of the magnetic field. It is noted from eqn. (11), that the parameter $\frac{2r^{2}}{w_{m}^{2}}$ essentially represents the magnetic field parameter $\alpha$ $\frac{2r^{2}}{w_{m}^{2}} = \frac{Br^{2}}{2} = \frac{\phi}{2\pi} = \alpha$ (e = $\hbar$ = 1). Actually, the external magnetic field modifies the Berry curvature and hence the phase. We can compute the effect of this magnetic field on the Berry phase through the evaluation of the modified OAM given by $\vec{R}\times \vec{P}$ with $\vec{R} = \vec{r} - \vec{A}(p)$ \cite{13} where the SU(2) gauge field $\vec{A}(p)$ represents the spin degrees of freedom and covariant momentum $\vec{P} = \vec{p} - \vec{A}(r),$ incorporates the gauge $\vec{A}(r)$ corresponding to the external field. Utilizing the same methodology used in the previous section, the SOI in presence of a magnetic field can be obtained from the modified OAM operator, which can be written as 
   \begin{eqnarray}
    \vec{\tilde{L}} &=& \vec{R} \times \vec{P}\nonumber\\
    &=& \left(\vec{r} - \vec{{\cal {A}}}(p)\right) \times \left(\vec{p} - \vec{A}(r)\right)\nonumber\\
    &=& \vec{r}\times \vec{p} - \vec{{\cal {A}}}(p)\times \vec{p} - \vec{r}\times \vec{A}(r) +  \vec{{\cal {A}}}(p)\times \vec{A}(r)\nonumber\\
    &=& \vec{L} - \vec{L}_{1} - \vec{L}_{2} + \vec{L}_{3}.
   \end{eqnarray}
   To find $\vec{\tilde{L}}$ we have to explicitly calculate $\langle \vec{L}_{1}\rangle, \langle \vec{L}_{2}\rangle$ and $\langle \vec{L}_{3}\rangle.$
   As the SU(2) gauge field $\vec{{\cal A}}(p) $ \cite{6} can be written as \begin{eqnarray} {\cal \vec{A}}(\vec{p}) =\mu \frac{\vec{p}\times \vec{\sigma}}{p^{2}} \label{23}\end{eqnarray},
   where $\vec{\sigma}$ is the vector of Pauli matrices,  we have
\begin{eqnarray}\langle \vec{L}_{1}\rangle = \langle \mu \frac{\vec{p}\times \vec{\sigma}\times\vec{p}}{p^{2}} \rangle = -\langle \mu\vec{p}\times(\frac{\vec{p}}{p^{2}}\times\vec{\sigma}) \rangle.\end{eqnarray} The expectation value of $\vec{\sigma}$ is given by
\begin{equation}\langle\vec{\sigma}\rangle = \frac{\langle\psi|\vec{\sigma}|\psi\rangle}{\langle\psi|\psi\rangle}\label{26},\end{equation}
   where $\psi$ is a two-component spinor
  \begin{equation} \psi = \left(\begin{array}{cr}
      \psi_{1} \\
       \psi_{2}
    \end{array}\right) \label{27}\end{equation} with $\langle\psi|\psi\rangle = 1.$  This gives \begin{eqnarray} \langle\vec{\sigma}\rangle = \vec{n}\label{28},\end{eqnarray}
     $\vec{n}$ being a unit vector. Thus, 
 \begin{eqnarray} \langle\vec{L}_{1}\rangle = -\langle \mu\vec{p}\times(\frac{\vec{p}}{p^{2}}\times\vec{\sigma})\rangle = -\langle\mu \vec{\kappa}\times (\vec{\kappa}\times \vec{\sigma})\rangle\label{25}\end{eqnarray} with $\frac{\vec{p}}{p} = \vec{\kappa},$ $\vec{\kappa}$ being the unit vector. This gives  \begin{eqnarray} \langle\vec{L}_{1}\rangle = -\mu \vec{n}\label{29}.\end{eqnarray} Similarly for $\vec{L}_{2}$ we have
   \begin{eqnarray}
   \vec{L}_{2} &=& \vec{r}\times \vec{A}(r)\nonumber\\
   &=&  \vec{r}\times \frac{\alpha}{r}\hat{\vec{e}}_{\phi} \nonumber\\
   &=& -\alpha\vec{\tilde{\kappa}}\times \hat{\vec{e}}_{\phi},
   \end{eqnarray}
   where $\vec{\tilde{\kappa}} = \frac{\vec{r}}{r}.$ This gives \begin{eqnarray}\langle\vec{L}_{2}\rangle = -\langle\alpha\rangle\vec{\tilde{n}}, \end{eqnarray}
   $\vec{\tilde{n}}$ being the unit vector.
   Now for the term $\vec{L}_{3},$ we write,
   \begin{eqnarray}
   \vec{L}_{3} &=&  \vec{{\cal {A}}}(p)\times \vec{A}(r)\nonumber\\
   &=& \mu\frac{\vec{p}}{p^{2}}\times\vec{\sigma}\times\alpha\frac{\vec{r}}{r^{2}}~~~~~~ (~with ~~\vec{r}~~ =~~r\hat{\vec{e}}_{\phi})\nonumber\\
   &=& \frac{\alpha\mu}{p^{2}r^{2}}\vec{p}\times \vec{\sigma}\times \vec{r}\nonumber\\
   &=& \frac{\alpha\mu}{p^{2}r^{2}} \vec{\sigma}\times\vec{r} \times\vec{p}\nonumber\\
   &=& 0.
   \end{eqnarray}
 As in this case the spin vector is orthogonal to the orbital momentum the expression $\vec{\sigma}\times \vec{L} = ~0,$ which is applied in the calculation of $\vec{L}_{3}$.
   
    The expected value $\alpha$ is given by \begin{equation} \langle \alpha\rangle = \left\langle \frac{2r^{2}}{w_{m}^{2}}\right\rangle = \frac{\left\langle\psi |\frac{2r^{2}}{w_{m}^{2}}|\psi\right\rangle}{\left\langle\psi |\psi\right\rangle}.\end{equation}
  We can write the solution of the above Schroedinger eqn. as 
   \begin{equation} \psi^{L}_{l,n} \simeq \left(\frac{r}{w_{m}}\right)^{|l|}L_{n}^{|l|}\left(\frac{2r^{2}}{w_{m}^{2}}\right)exp\left(-\frac{r^{2}}{w_{m}^{2}}\right)exp\left[i(l\phi+k_{z}z)\right].\label{6}\end{equation}
   This gives \cite{14}
   \begin{equation}
   \langle \alpha\rangle = \langle \frac{2r^{2}}{w_{m}^{2}}\rangle = 2n + |l| + 1,\end{equation}
   where $n$ is the radial quantum number.
   Thus we have \cite{sub}
   \begin{eqnarray}
   \langle\vec{\tilde{L}}\rangle &=& l+ \langle \vec{L}_{1}\rangle + \langle \vec{L}_{2}\rangle \nonumber\\
   &=& l + \mu + g~ \langle \alpha\rangle \nonumber\\
   &=& l + \mu + g~ (2n + |l| + 1),
   \end{eqnarray}
   where $g$ denotes the sign of the external magnetic field.
   This also suggests a SOI effect as in case of free EVB. The modified Berry curvature of the electron vortex beams is then given by  
    \begin{equation}\tilde{\vec{\Omega}}(\vec{k}) = (\mu + g\langle\alpha\rangle) \frac{\vec{k}}{k^{3}},\label{44}\end{equation}
    where $g\langle\alpha\rangle \frac{\vec{k}}{k^{3}}$ is the extra modification term in the Berry curvature due to external magnetic field and $g = \pm 1$ depending on the orientation of the magnetic field.  
    The  modified anomalous velocity is then given by
    \begin{equation} \vec{\tilde v}_{a} = \left(\mu + g\langle\alpha\rangle\right) \frac{\dot{\vec{k}}\times\vec{k}}{k^{3}}.\end{equation} Now from eqn (25) we note that the expected value of $\langle\alpha\rangle$ is given by $(2n + |l| + 1)$ where $n$ is the radial quantum number, ($ n = 0, 1, 2$......) and $|l|$ is the orbital angular momentum carried by the beam. 
 In terms of the above form of $\langle\alpha\rangle$ we can rewrite the modified anomalous velocity as
     \begin{equation}\vec{\tilde{v}}_{a} = \left(\mu + g\langle\alpha\rangle\right) \frac{\dot{\vec{k}}\times\vec{k}}{k^{3}} = \left[\mu + g(2n + |l| + 1)\right] \frac{\dot{\vec{k}}\times\vec{k}}{k^{3}},\end{equation} with $\mu = \pm\frac{1}{2}$ in the local non-inertial frame and $n$($|l|$) is an integer. 
   The quantized value of $\mu=\pm \frac{1}{2}$ corresponding to the spin given by $|\mu|$ with $s_z=\pm \frac{1}{2},$ we note that the modified value of the Berry phase factor $|\mu+g\langle\alpha\rangle|$ corresponds to the total spin of the system. In our case the corresponding vortex is the spin vortex, the factor $\langle\alpha\rangle = (2n + |l| + 1)$ will give rise to the clustering of spin $\frac{1}{2}$ states. The resultant spin of the cluster is given by \cite{sub} $|\mu + g\langle\alpha\rangle| =| \mu + g (2n + |l|+1)|.$  It is noted that for positive $\langle\alpha\rangle$ the minimum value of $\langle\alpha\rangle = ~1$ when n($|l|$) is taken to be zero.
 This gives rise to the helicity  states for $\mu = \frac{1}{2}$ (-$\frac{1}{2}$), $\mu + \langle\alpha\rangle =~ \frac{3}{2}$ ($\frac{1}{2}.$) For higher values of $n$($|l|$), we will have higher half integer values of ($\mu + \langle\alpha\rangle$).
   One point may be noted that for $+\langle\alpha\rangle$ we will always have up spins (say)and conversely negative $\langle\alpha\rangle$ corresponds to the down spins. The orientation of the magnetic field thus determines whether we achieve positive spin current or negative spin current generating in the output. This leads to the spin filtering \cite{sub}. This situation is depicted in the fig \ref{b}.
   
 Now if we can manage to have an external field in such a way that the orientation of the field is altered in two consecutive time sequences, we have alternating spin Hall current in consecutive time sequences in opposite directions.
   
Thus if an external time dependent magnetic field is introduced in the longitudinal direction, we can tune the spin Hall current in case of a tilted  electron vortex beams. This gives rise to spin filtering as well. From eqn. (10), we can have enhanced spin current in this case. 
Spin current is enhanced due to the presence of the external magnetic field.
 \begin{figure}\label{b}
   \includegraphics[width=10.0 cm]{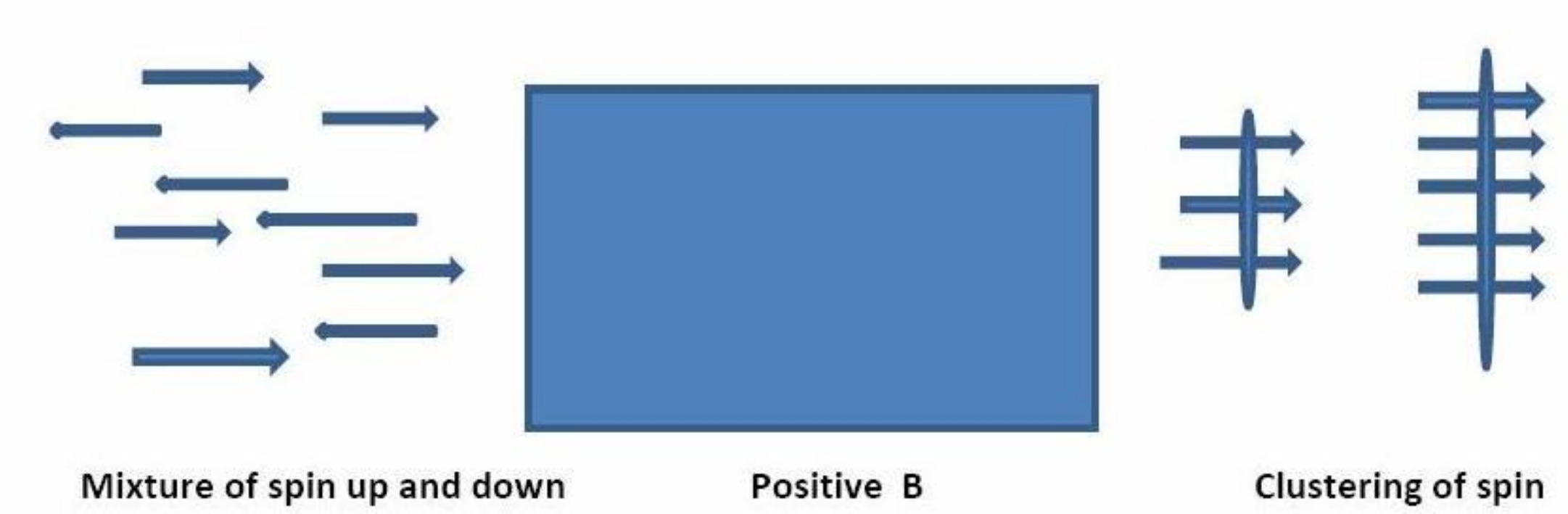}
   \hspace{1cm}
   \caption{ Spin filter and clusturing of spins.}
   \end{figure}
   
Thus spin filtering occurs and either positive or negative spin Hall currents appear depending on the orientation of the magnetic field with clustering of spins.
For a hopping field we can successfully achieve alternating spin Hall currents with positive and negative spin states. This is one of the main result in this paper.
\section{Discussion}
In the quantization scheme of a fermion, the vortex beam is a natural consequence of the skyrmion model of an electron. In this situation a fermion can be thought of as a scalar particle encircling a vortex line which is topologically equivalent to a magnetic flux line. The geometric phase acquired by the scalar particle moving around the vortex line in a closed loop essentially gives rise to the specific properties of a fermion such as the sign change of the wave function after a $2\pi$ rotation as well as the spin-statistics relation. Interestingly, when the vortex line is tilted with respect to the propagation direction of the wave front, the related Berry phase involves non-quantized monopole. From an analysis of the RG flow of the monopole charge, it is argued that in this case we have temporal variation of the direction of the vortex line giving rise to spin Hall effect which can be observed experimentally.

Also we have encountered the situation of applying an external magnetic field on the freely propagating electron vortex beams. In the present study we show that in presence of an external magnetic field in time space the Berry curvature is changed and the phase is modified by the magnetic field parameter. This parameter is associated with the Gouy phase which is related to the diffractive LG beams in free space. In a magnetic field, the Gouy phase factor is associated with the transverse kinetic energy of spatially confined mode and  and leads to the contribution of the Gouy energy to Landau energy. Indeed, this parameter determines the squared 'spot size' of the LG beam. It has been argued here that when electron vortex beams propagate in an external time dependent magnetic field we have modification of the spin Hall current from that in the free space for tilted vortex. It is found that this leads to spin filtering when either positive or negative spin Hall currents appear depending on the orientation of the magnetic field with clustering of spins. For a hopping field which alters its orientation at two consecutive time sequences, we have alternating spin Hall currents with positive and negative spin states at successive time sequences.




\begin{center}
{\bf Acknowledgments}: We would like to acknowledge Prof. P. Bandyopadhyay for fruitful discussions.
\end{center}

\end{document}